\documentclass[twocolumn,showpacs,preprintnumbers,amsmath,amssymb]{revtex4}
\usepackage{tabularx,graphicx}

\usepackage{color}
\usepackage{hyperref}
\hypersetup{
    colorlinks=true,
    linkcolor=blue,
    filecolor=blue,      
    urlcolor=blue,
}

\usepackage{color}

\newcommand{\bluee}{\textcolor{black}}

\usepackage{ulem}   

\begin{document}

\newcommand{\beq}{\begin{equation}}
\newcommand{\eeq}{\end{equation}}
\newcommand{\beqn}{\begin{eqnarray}}
\newcommand{\eeqn}{\end{eqnarray}}
\newcommand{\bmath}{\begin{subequations}}
\newcommand{\emath}{\end{subequations}}
\newcommand{\bra}[1]{\langle #1|}
\newcommand{\ket}[1]{|#1\rangle}


  \title{Reply to ``Is $MgB_2$ a superconductor? Comment on “Evidence Against Superconductivity in Flux Trapping Experiments on Hydrides
Under High Pressure''''} 

\author{J. E. Hirsch$^{a}$  \footnote{Corresponding author: jhirsch@ucsd.edu}  and F. Marsiglio$^{b}$  \footnote{email: fm3@ualberta.ca}  }
\address{$^{a}$Department of Physics, University of California, San Diego,
La Jolla, CA 92093-0319\\
$^{b}$Department of Physics, University of Alberta, Edmonton,
Alberta, Canada T6G 2E1}

\begin{abstract} 
The preceding Comment \cite{comment}, previously posted as arXiv:2312.04495 \cite{talantsevarxiv}, 
on our paper J. Supercond. Nov. Mag. 35, 3141 (2022) \cite{hmtrapping2022}
provides a welcome opportunity to clarify what we understand to be pervading misconceptions 
by  Eremets, Minkov  and coauthors in regard to our analysis \cite{hmtrapping2022} of their trapped flux experiments in 
hydrides under pressure \cite{minkovtrappedpub}.
We hope that  this Reply \cite{ourreply}  will help   readers interested in hydride superconductivity sort out between different claims and counterclaims 
in the literature and inform their views based on verifiable facts.
\end{abstract}

\maketitle 

\section{ introduction}
In their Comment \cite{comment,talantsevarxiv}, the authors make some  strong claims about our paper Ref. \cite{hmtrapping2022} which, if factual,
would call into question our scientific conclusions on their work on trapped flux in hydrides under
pressure  \cite{minkovtrappedpub},  as well as raise serious questions on our scientific idoneity and our scientific integrity. In particular, quoting verbatim from Ref. \cite{comment}  commenting on our 2022 paper Ref. \cite{hmtrapping2022} (in the following, reference $^1$ refers to Ref. \cite{hmtrapping2022} here, and ``the authors'' are
JEH and FM):
\begin{enumerate}
  \item {\it ``relies on the wrong model coupled with selective manipulations
(hide/delete) of calculated datasets''}
  \item {\it ``ignores the reference measurements after the release of
pressure''}
  \item {\it ``the authors have restricted the publication of their simulation code, which
we view as a breach of scientific integrity and open science principles.''}
  \item{\it ``the authors$^1$ did not verify their model using any ZFC or FC data
measured in well-studied superconductors.''}
  
      \item {\it ``Upon examination of the provided code, we surprisingly discovered that the code$^1$ is a
simulation tool''}
  \item {\it ``Significantly, the Hirsch-Marsiglio model$^1$ does not describe the Meissner regime
  ...Instead, the model predicts a significant positive magnetic moment for ZFC $m(B_{appl}=0,T)$ ... This feature of the proposed model was not discussed by the authors$^1$ and obviously contradicts
the physics for the trapped magnetic flux in superconductors.''}

    \item {\it ``the authors$^1$ in their Figures 3-5 hid/deleted (without
reporting this) parts of their simulated datasets that disagree with the Meissner state. This potentially
allowed them to conceal the issue that their model and computer code do not adequately describe the
Meissner state.''}

\item   {\it ``the authors$^1$ engaged in methodological
malpractice by creating a FORTRAN 77 simulation code rather than performing a genuine data fit as
claimed''}

  \item {\it``We
argue that the authors overlooked reference measurements conducted on evidently non-superconducting
samples, which demonstrate the absence of such artefacts stemming from the sample environment
(diamond anvil cells, rhenium gaskets, etc.). The reference measurements revealed that the temperature-dependent
magnetic moment of the same samples within the same diamond anvil cells, after the release
of pressure, does not exhibit the trapping of magnetic flux.''}
  \item{\it ``The simulations conducted by the
authors$^1$ involve unverified models, fixing fitting parameters without proper argumentation, and
unjustifiably deleting parts of the simulation dataset.''}
  
  \end{enumerate}

These  claims have previously been published by the authors in arxiv Ref. \cite{talantsevarxiv}.
  We will show that all of them are unfounded.
  
  \section{ relevant literature}
 In fact, the  information  that invalidates every one of the claims listed in Sect. I already exist in the scientific literature. 
  Here we list the relevant papers that the reader can consult to verify this. In the following section 
  we discuss the claims in detail to make this Reply self-contained.
  
  1) The original report of the trapped field measurements by Minkov et al. was in the arxiv posting
  arXiv:2206.14108v1 of June 28, 2022, hereafter Ref. \cite{minkovarxivv1}.
  Our paper being commented on, Ref. \cite{hmtrapping2022}, was written in response to that posting and published
  on-line August 10, 2022.
Subsequently the authors updated their arxiv posting with   arXiv:2206.14108v2 on September 25, 2022 \cite{minkovarxivv2},
that updated version was then published in the journal Nature Physics \cite{minkovtrappedpub}.

2) On December 7, 2023, the authors of this Comment posted a paper 
arXiv:2312.04495v1, essentially identical to this Comment, hereafter Ref. \cite{talantsevarxivv1}.
It was later updated with minor changes as Ref. \cite{talantsevarxiv}.

3) On December 8, 2023, we sent a letter to the authors of Ref. \cite{talantsevarxivv1} explaining
why their posting  Ref. \cite{talantsevarxivv1} was wrong and misleading to the scientific community.
The authors did not acknowledge receipt of our letter nor did they take corrective action. 
Our letter is reproduced in the Appendix of this Reply.

4) On December 18, 2023 and January 15, 2024, one of us (JEH) posted responses to some of    the authors' claims in \cite{talantsevarxiv},
which are identical to  claims in this Comment \cite{comment}, at OSF \cite{osfmyresponse} and arxiv \cite{arxivmyresponse}
respectively.
In this Comment \cite{comment} the authors neither cite nor acknowledge the existence of  these responses
\cite{osfmyresponse,arxivmyresponse} to their claims,  nor do they
acknowledge receipt of our letter of December 8, 2023 (item 3) here and Appendix) addressing those claims.

5) On April 12, 2024, we published Ref. \cite{further}, where we presented an analysis of flux trapping experiments on
known superconductors reported by Bud'ko, Xu and Canfield on September 13, 2023 \cite{canfield} using the same ``Hirsch-Marsiglio model'' used in our
paper Ref. \cite{hmtrapping2022}, provided further analysis of the  flux
trapping experiments in hydrides reported in \cite{minkovarxivv1,minkovarxivv2,minkovtrappedpub}, and
responded to the criticism in \cite{talantsevarxiv}, which is  essentially the same as this Comment \cite{comment}. In this Comment \cite{comment} the authors neither cite nor acknowledge the existence of our Ref. \cite{further} nor in particular   our response to their claims presented in Sect. 5 of Ref. \cite{further}.

6) On May 26, 2024, we posted on arxiv Ref.  \cite{tinyarxiv}, as well as submitted  for publication, our analysis
of a recent paper of Bud'ko et al. on trapped flux in tiny samples of known superconductors under pressure \cite{tinybudko} and its relation with the trapped
flux experiments on hydrides \cite{minkovtrappedpub}, using the same
so-called \cite{comment}  ``Hirsch-Marsiglio model'' being criticized in this Comment \cite{comment}. 
In this Comment \cite{comment} the authors neither cite nor acknowledge the existence of  this paper \cite{tinyarxiv}
 nor address its relevance to their claims.
 
 In summary, subsequent to our paper \cite{hmtrapping2022} being commented on in the authors' Comment Ref. \cite{comment},
 we published four papers in the scientific literature, namely Refs. \cite{ osfmyresponse,arxivmyresponse,further,tinyarxiv},
 using the same model that we used in  \cite{hmtrapping2022}, that are relevant to the authors' Comment but are not
 cited nor discussed in their Comment \cite{comment}. 
 
 \section{refutation of the authors' claims}
 We refute here the authors' claims 1. to 10. listed in the Introduction.
 
 \subsection{Claim that we hid/deleted data} 
 In their first version of their report on experimental results for flux trapping in hydrides \cite{talantsevarxivv1}, the authors
 presented a theoretical analysis of their data 
  {\it ``in terms of the classical critical state Bean model''}, for which they cited Refs. \cite{bean1,bean2}. Their Eq  (2), 
  claimed to represent the prediction of that model, was:
\beqn
m_{trap}(H_M)&=&\int_{r}^{d/2} \pi r^2 dI \nonumber \\
 &=&m_s[1-(1-\frac{H_M-H_p}{2H^*-H_p})^3] .
\eeqn
They stated after writing this equation that {\it ``radius of penetration of circulating currents, $r$, varies between $d/2$ (no trapped
flux) and zero (at saturation) and is considered in a linear approximation as 
$\frac{d}{2}(1-\frac{H_M-H_p}{2H^*-H_p})^3$''}, and  that they applied the boundary condition 
{\it ``$m_{trap}=0$ if $H_M \le H_p$''}. 

The authors then showed in their Fig. 2 of  \cite{talantsevarxivv1}   that their 
Eq. (2) (Eq. (1) here) {\it ``provides a reasonable fit of the experimental data illustrated in Figure 2.''}.
Those data were {for the ZFC measurement  protocol}, namely the procedure where the sample is first cooled, then the 
magnetic field $H_M$ is applied at low temperatures, then the external magnetic field is removed and the remnant
magnetic moment is measured at the same low temperature. Note that Eq. (1) predicts that $m_{trap}(H_M)$ is 
{\it linear} in $(H_M-H_p)$ for small $(H_M-H_p)$.

In our paper being commented on, Ref. \cite{hmtrapping2022}, we stated that the theoretical fit performed by the authors
of Ref. \cite{talantsevarxivv1} (Eq. (1) above) assumed the trapped magnetic moment was given by
\beq
m=\int_{r}^{d/2} \pi r'^2 j_c h dr'=m_s[1-(\frac{r}{d/2})^3]
\eeq
\beq
r=r(H_M)=\frac{d}{2}(1-\frac{H_M-H_p}{2H^*})
\eeq
which are of course identical to Eq. (1) above. 
We explained that $H_p$ in the ZFC protocol is the {\it ``threshold value of the applied field where it
begins to penetrate the sample at low temperatures''}. We also stated right thereafter 
{\it ``so that $r(H_p)=d/2$''}. We did not state explicitly that  Eq. (1) or the equivalent Eqs. (2,3) above 
should NOT be used for $H_M<H_p$, but this was completely obvious from the context:
for $H_M<H_p$ Eq. (3) yields a radius $r(H_M)$ larger than half the diameter $d/2$, which of course makes no sense, and would
be completely extraneous to the Bean model, on which these equations are based.  
As the authors themselves had stated, {\it ``radius of
penetration of circulating currents, r, varies between d/2
(no trapped flux) and zero (at saturation)''}. We assumed that the boundary condition $m_{trap}=0$ if $H_M \le H_p$, which had been explicitly stated in
Ref. \cite{talantsevarxivv1}, would be obvious to readers and didn't state it again explicitly.

Then we proceeded to argue in Ref. \cite{hmtrapping2022}  that Eqs. (2), (3) (and hence Eq. (1)) are not appropriate for the ZFC protocol within the Bean model, 
because they imply that when the field is applied at low temperatures  it fully penetrates, which is inconsistent with the Bean model, 
but instead are appropriate to model a FC protocol, where the field fully penetrates at a high temperature, before
the system is cooled in the presence of the field and the external field is subsequently removed. And that Eqs, (2), (3)  would be appropriate to model 
FC data (that did not exist at the time) provided $H_p$ in Eq. (3) is set to 0 and $2H^*$ in Eq. (3) is replaced
by $H^*$. And, that the correct way to apply the Bean model to a ZFC protocol is to use, instead of Eqs.  (2) and (3),
\beq
m=m_s[1- 2 (\frac{r_1}{d/2})^3+(\frac{r_2}{d/2})^3]
\eeq
where, for $H_M<H^*+H_p$
\bmath
\beq
r_1=\frac{d}{2}(1-\frac{H_M-H_p}{2H^*})
\eeq
\beq
r_2=\frac{d}{2}(1-\frac{H_M-H_p}{H^*}),
\eeq
\emath
and showed in Fig. 2 of \cite{hmtrapping2022} the physical meaning of $r_1$ and $r_2$.
We did not  state explicitly that Eqs. (5) did not apply
to the region $H_M<H_p$, which would yield values of $r_1$ and $r_2$ $larger$
than the radius of the sample $d/2$, because it was completely obvious from
the context, from the Bean model,  and from the diagram in Fig. 2  of our
paper \cite{hmtrapping2022} what the physical meaning
of $r_1$ and $r_2$ was, both $smaller$ than the  radius  of the sample $d/2$.
The key difference between Eqs. (4), (5) and Eqs. (2), (3) is that Eqs. (4), (5) predict 
{\it quadratic} dependence of $m$ versus $(H_M-H_p)$ for small  $(H_M-H_p)$ instead of linear as Eqs. (2), (3) do.

Subsequently in Ref. \cite{hmtrapping2022}, we used our model Eqs. (4), (5) to show what ZFC data should
look like for a variety of model parameters, stressing in particular that the trapped moment
under ZFC goes to zero quadratically as $H_M$ approaches $H_p$ from above, 
and not linearly as the authors' experimental data and their own linear fit to them showed \cite{minkovarxivv1}.
In all of the examples we showed (Figs. 3, 4, and 5 of Ref. \cite{hmtrapping2022}) we 
showed calculated ZFC values as dashed lines in red, and in all the figures
the dashed red lines were plotted only for $H_M \ge H_p$, and there are no
red lines for $H_M<H_p$, because of course the model says that
if the applied field $H_M$ is smaller than the threshold value $H_p$ where the field starts
to penetrate the sample, no trapped moment will result when the applied field is
removed.

In subsequent publications \cite{further,tinyarxiv} we applied our model to the data 
for trapped flux reported recently by Bud'ko and coauthors for known superconductors \cite{canfield},
and showed that in all cases the measured behavior for ZFC was consistent with the behavior
predicted by our model, namely quadratic   $H_M-H_p$, contrary to what was reported
for hydrides \cite{minkovtrappedpub}. In all cases we only plotted moments for $H_M>H_p$ 
because there can be no trapped moment if the field is not large enough to enter the sample.

We hope that readers will understand, based on the above, that our model
{\it does not} predict a trapped magnetic moment for $H_M<H_p$. 
Yet, the authors of the Comment \cite{comment}  proceeded to use our Eqs. (4), (5) to plot magnetic moment
versus field $including$ what results from these equations in  the region $H_M<H_p$ in their Figs. 
1 (a), 1 (b), 2 (b) and 2 (c),  and claim (see 1., 6., 7., 10. in the Introduction of this paper) that our
model does not describe 
the Meissner regime $H_M<H_p$ because it predicts a finite moment in that region and that we ``hid/deleted''
the data predicted by our model in that region.

This is particularly unfathomable in view of the fact that we wrote to the
authors long ago explaining their misuderstanding (see Appendix)
and that we published four papers after Ref. \cite{hmtrapping2022},
namely \cite{osfmyresponse,arxivmyresponse,further,tinyarxiv}, addressing the same issue again and again.
 
\subsection{Other claims} 

Regarding the claim 2. in the Introduction that our model ignores the reference measurements
after the release of pressure, and the equivalent claim 9.,
we point out that we responded to it in Sect. 5 of Ref. \cite{further},
which we repeat here verbatim:
\newline
\newline
\noindent {\bf ``Does our paper ignore the reference measurements after the release of pressure?}
In Refs. \cite{minkovarxivv1,minkovtrappedpub} it is stated that after a diamond cracked and pressure dropped below
10GPa, no trapped moment was detected in the experiment. 
In our paper we hypothesized that the measured moments, which were all measured  with uncracked diamonds and under much higher pressures, originated from 
{\it ``magnetic properties of the sample or its environment unrelated to superconductivity''}. 
The authors of \cite{talantsevarxiv} object to the fact that we did not consider the measurements with cracked diamond and low
pressure relevant. We do not, because 
both the properties of the sample and its environment will change when diamonds crack and the pressure drops,
so we don't find it surprising nor relevant to the understanding that no moments were measured under those
very different experimental conditions. 

The authors should perform control experiments, under the same experimental conditions
for which moments were detected (in particular high pressures and uncracked diamonds), using samples known not  to be superconducting, as well as samples known to be
superconducting,
to provide relevant information to support or refute their claims \cite{minkovarxivv1,minkovtrappedpub} 
 that their trapped flux experiments
shed light on the question whether the hydrides are superconducting or not.''

Regarding their claim 4., that we didn't verify our model using known
superconductors, we point out that (a) the Bean model on which our model is based
is a well known and generally
accepted model and (b) we did exactly that in our 
papers \cite{further,tinyarxiv}.

Regarding their claims 5., 8. and 10.,  that our model is  a ``simulation code''
rather than a ``genuine data fix'' and that this constitutes ``methodological malpractice'', we responded to it in Sect. 5 of Ref. \cite{further},
which we repeat here verbatim:

``In our calculations both in Ref. \cite{hmtrapping2022}
and in other subsequent papers \cite{further, tinyarxiv}  we varied the parameters to obtain good fits to the data. Our model results vary smoothly with the
parameters and each parameter influences the resulting curve in a distinct way, so it is simple to find parameters that will
best represent the measured data.~'' Then we explained with an example in Fig.~3 of that paper
how that works, as follows: ``As an example consider the three red curves shown in the upper left panel of
Fig.~3. It is clear that for other values of $H^*$
\bluee{ and the same value of $H_p$}
 the curves will interpolate between the ones given and that
no value of $H^*$ will give a good fit to the data.''

Regarding their claim 3.  about us restricting publication of our simulation
code, we discuss it
in the following section.

 \section{relevant history and accusations by the authors regarding computer code}
 Here we recount some historical events that are relevant to understanding 
 some of the statements in the authors' Comment \cite{comment}.
 
 In June 2023, several of the authors of the Comment \cite{comment}, including M. I. Eremets and V. Ksenofontov,
 as well as we, the authors of this Reply, attended the meeting 
 Supertripes 2023. During that meeting, we discussed with Eremets and particularly with Ksenofontov our paper
 Ref.~\cite{hmtrapping2022}. Ksenofontov said that he wanted to understand how the curves in our figures were obtained,
 and we pointed out that they are simply calculated from the equations in the paper, given above. Nevertheless, he said 
 he would appreciate if we would send him our computer codes.
 
 On September 18, 2023, Ksenofontov wrote to us asking for a  code, stating that "I am confident that together we will reach a consensus". On September 19, one of us (JEH) sent him one of his Fortran codes. We did not include instructions on how to use and interpret what the code printed out, expecting he would contact us if any questions arose. The version of the code that was sent printed out also numbers for values of field $H_M<H_p$ in Eqs (4,5) that have no physical significance and should be ignored. 
 Instead, in their Dec. 8 2023 arxiv posting \cite{talantsevarxivv1} and in this Comment \cite{comment} the authors plot red
 curves including the region $H_M<H_p$ and claim that those are ``predictions'' of our model. This is despite the fact 
 that we informed them immediately after their arxiv posting \cite{talantsevarxivv1} (see Appendix) that our model, just as 
 Bean's model, predicts no trapped moment for $H_M<H_p$.
 
 In \cite{talantsevarxivv1}  the authors also posted our computer code. They did this without our permission, which is in violation of intellectual property principles.
 
The authors of \cite{comment}  accuse us of restricting the publication of our computer code 
 in their Comment and state that this is a breach of scientific
 integrity. Our computer code simply codifies the equations given in the paper, which anybody can reproduce. It is not a breach of scientific integrity to deny permission to the authors to publish our code in their Comment, given that they continue to misrepresent  
 what the code calculates,  as explained above. We shared our code with those authors in good faith, upon their request, and that does not entitle them to publish our code in their papers to misinterpret it and malign it, as they did in  \cite{talantsevarxivv1},  which they still have not retracted.
 
 Finally, to claim that we  {\it ``engaged in methodological malpractice by creating a FORTRAN 77 simulation code rather than performing a genuine data fit as claimed''} is completely wrong. The particular 24-line computer code that we sent them is one of many computer codes that we used to simulate, fit and analyze the predictions resulting from our model in Ref.~\cite{hmtrapping2022}
 and their relation with the measured data.

\section{other relevant papers by these authors and other authors}

According to what was reported in the magnetic moment measurements on the same hydride samples in Ref. \cite{e2021p,e2021pcorr}, the field $H_p$ for $H_3S$  for which the magnetic moment versus field starts to deviate from linearity
indicating that the field starts to penetrate the sample, is approximately 95 mT at low temperatures.
In their revaluation analysis \cite{revaluation} the author estimated it to be even higher,
namely 108 mT. Instead, from their trapped magnetic flux measurements the authors deduced 
the threshold value of field for which the same samples trap magnetic flux to be $42 mT$  \cite{minkovtrappedpub}. 
It is difficult to understand how a sample that excludes magnetic fields smaller than 108mT could trap
magnetic fields between 42 mT and 108mT. 
According to the authors of Ref.  \cite{comment} themselves, such behavior 
{\it ``obviously contradicts
the physics for the trapped magnetic flux in superconductors.''}

In this context it is also relevant to note that Ref. \cite{comment} cites Ref. \cite{tallon} in support of their work.
In Ref. \cite{tallon},  Tallon argues that the ZFC trapping data reported in \cite{minkovtrappedpub} are
consistent with the $quadratic$ field dependence expected for such measurements that we pointed out \cite{hmtrapping2022} if
the threshold field is taken to be $H_p=24 mT$. 
Such a low estimate for $H_p$ is inconsistent with what the authors of Ref.  \cite{minkovtrappedpub} 
themselves estimated (42 mT)  and even more so from what results from their measurements in Refs. \cite{e2021p} and  \cite{revaluation},
as discussed above.


The authors of Ref. \cite{comment} also cite Ref. \cite{prozorov} in support of their work.
In Ref. \cite{prozorov}  Prozorov of course agrees with us \cite{hmtrapping2022} that the Bean model predicts
quadratic behavior in field for ZFC, contrary to what the authors erroneously said in Ref. \cite{minkovarxivv1}. 
In addition he  points out that behavior different from the quadratic behavior predicted by the Bean model
is predicted by other models, and claims that the reported behavior for flux trapping in hydrides can be consistent
with other models. However he fails to explain why in all the other recent reports of flux trapping for known superconductors
the behavior is quadratic, as we pointed out in Refs. \cite{further,tinyarxiv}. 

       \begin{figure} [t]
 \resizebox{7cm}{!}{\includegraphics[width=6cm]{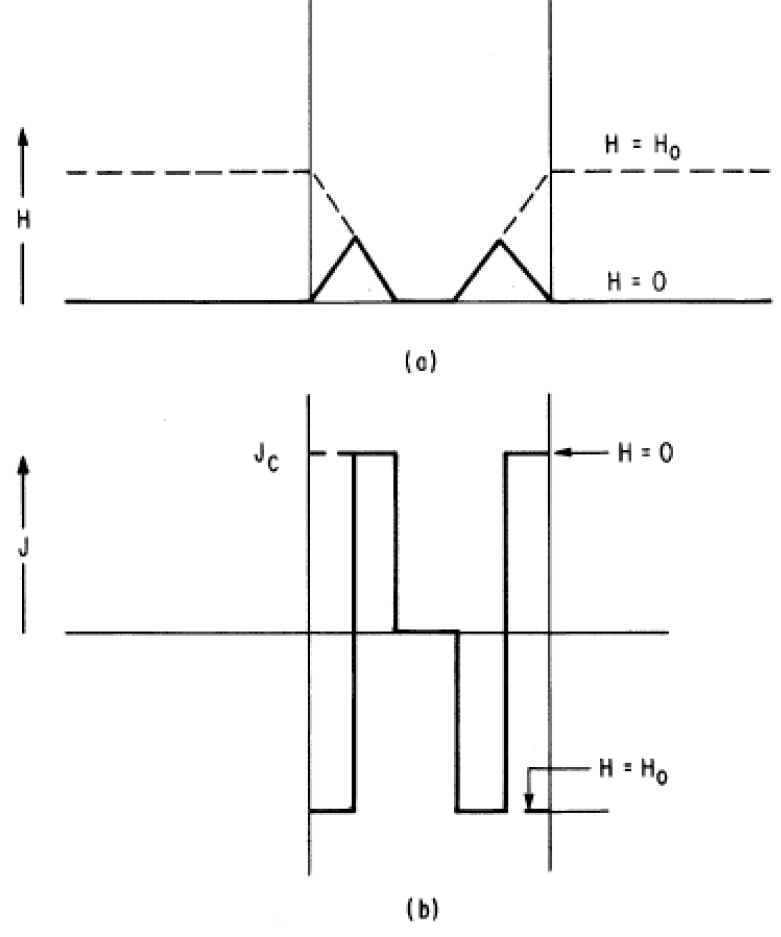}} 
 \caption {Figure 2 of Ref. \cite{bean2} by R. P. Bean.  Its caption reads
 {\it ``A plot of local fields and current density in a slab
after a field .$H_0$  has been applied and removed }}
 \label{figure1}
 \end{figure}

We also note that Ref. \cite{prozorov} 
points out that curves similar to those obtained by the authors for trapped moment \cite{minkovtrappedpub} would result
 {\it ``if similar measurements were performed on a ferromagnetic
sample with pinning. In this case, the remanent
magnetic moment is also positive and its field dependence is
due to the H-dependent size of the minor M (H) loops when
an applied magnetic field is insufficient to fully remagnetize
the sample (i.e. reorient all magnetic domains).''}, and for that reason 
{\it ``the demonstration of a diamagnetic response is imperative.''} to infer that the samples are superconducting.
Because that ``demonstration'' is supposed to be the data for magnetic moment versus field presented in Ref. \cite{e2021p} with an enormous
diamagnetic background subtraction \cite{e2021pcorr} and with the enormous discrepancy between
measured data and published data shown in  Ref. \cite{revaluationours},  we argue that 
it cannot be said that the conclusions of Ref.  \cite{prozorov} support the claim of the authors
that their trapped flux measurements demonstrate superconductivity.

Finally, we point out that the claim of the authors of Ref. \cite{minkovtrappedpub} that their measurements show flux creep 
indicating superconductivity was recently disproven  by Zen in Ref. \cite{zen},
further casting doubt on the claim that the trapped flux measurements indicate superconductivity.

\section{scientific ethics}

The issue of ``hiding/deleting'' data that the authors of the Comment 
keep bringing up \cite{comment,talantsevarxivv1,talantsevarxiv} should not be an issue.

As explained in our paper Ref. \cite{hmtrapping2022}, our model is  based on the Bean model.
Fig. 1 shows Fig. 2 of the Bean model from Ref. \cite{bean2}, for a slab geometry,
and Fig. 2 shows Fig. 2 of our paper Ref. \cite{hmtrapping2022} for a disk geometry.
The physics of the right panels of Fig. 2, corresponding to the ZFC protocol, is exactly the same
as the Bean physics shown in Fig. 1. Obviously the position where the field peaks in the interior
of the sample, which we called $r_1$, and the position where the field goes to zero in the interior of the
sample, which we called $r_2$, have to be inside the sample, whether it is a slab or a disk,
as  Fig. 1 (a)  and Fig. 2 center right panel    show. Obviously it makes no sense to consider a situation where
those positions are outside the sample, which is what results from Eqs. 5 if one uses them
for $H_M<H_p$. We should acknowledge however that, as correctly pointed out in the Comment \cite{comment},
the sign of the currents in the lower right panel of Fig. 2 should be reversed to be consistent with the
center right panel of Fig. 2,  as is shown correctly in Fig. 1(b). This minor point is however completely irrelevant to the issue at hand.

         \begin{figure} [t]
 \resizebox{7cm}{!}{\includegraphics[width=6cm]{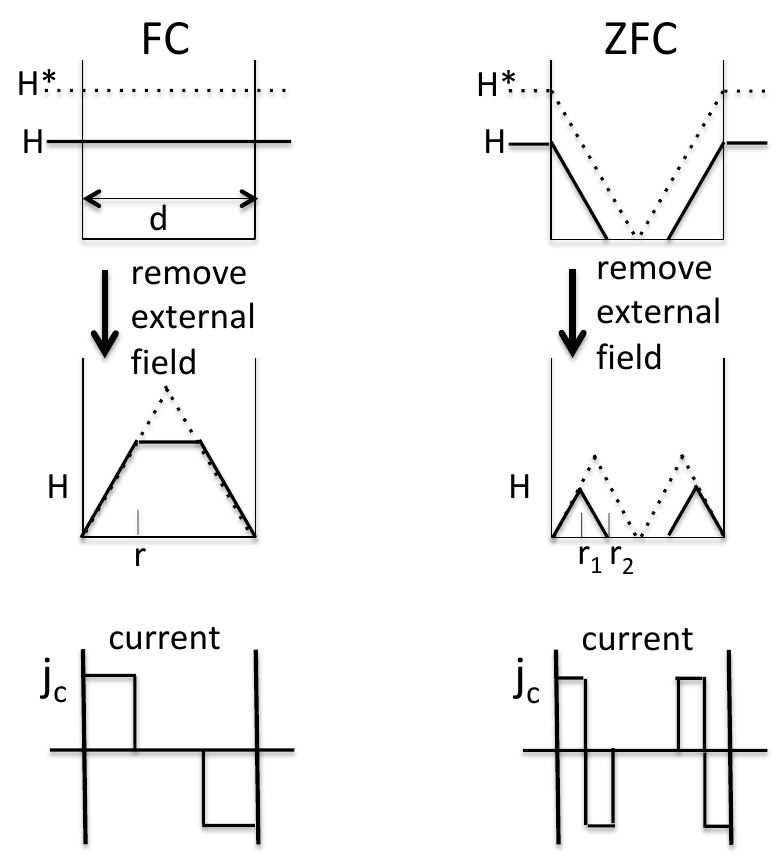}} 
 \caption {Figure 2 of our paper Ref. \cite{hmtrapping2022}. Its caption reads 
 {\it ``Magnetic fields and currents predicted by the Bean model
under field-cooled (FC) and zero zero-field-cooled (ZFC) protocols. Here
we assume $H_p=0$   for simplicity.''}   }
 \label{figure1}
 \end{figure}

The relevant unresolved issue  relates to the fact that the authors of the Comment \cite{comment} {\it must know}, given Figs. 1 and 2, 
the contents of our paper \cite{hmtrapping2022} and our subsequent papers
\cite{further,tinyarxiv}, Refs. \cite{osfmyresponse,arxivmyresponse}, and our letter to them reproduced in the Appendix,
that the collection of  statements in their Comment:
\newline
\newline
\noindent
{\it ``selective manipulations
(hide/delete) of calculated datasets... Significantly, the Hirsch-Marsiglio model$^1$ does not describe the Meissner regime:...
Instead, the model predicts a significant positive magnetic moment for ZFC $m(B_{appl}=0,T)$  
(see Figure 1):...This feature of the proposed model was not discussed by the authors$^1$ and obviously contradicts
the physics for the trapped magnetic flux in superconductors...manually hide/delete all simulated points within the Meissner state, i.e., for $\mu_0H_{appl}\le \mu_0 H_p$... plot the partially deleted simulated $m(\mu_0H_{appl})$ curve...
It is evident that for all $B_{appl}<\mu_0 H_p$,   the simulated curve does not show the Meissner state, namely
$m(B_{appl},T)=0$ for all $B_{appl}<\mu_0 H_p$...demonstrating the unphysical anomaly within the
Meissner state...we conclude that the authors$^1$ in their Figures 3-5$^1$  hid/deleted (without
reporting this) parts of their simulated datasets that disagree with the Meissner state. This potentially
allowed them to conceal the issue that their model and computer code do not adequately describe the
Meissner state...unjustifiably deleting parts of the simulation dataset"}.
\newline
\newline
\noindent together with  the portion of their red curves in their figs. 1 and 2 for $H_M<H_p$ attributed to our model, are utterly and blatantly {\bf false}.

Therefore, the question that needs to be answered is: {\it \textbf{Why do the authors of Ref. \cite{comment} choose to knowingly write such falsehoods and submit them for publication in a scientific journal?}}

We cannot conceive of an answer to that question that is consonant with ethical scientific behavior. That is why we are raising
the question here. Perhaps readers of the Comment \cite{comment} and this Reply  can, and if so hopefully will share it with the scientific community.

\section{discussion}
More than two  years has gone by since the authors of the Comment \cite{comment,talantsevarxiv} published their
measurements \cite{minkovarxivv1} showing the anomalous behavior (linear rather than quadratic) of trapped moment
versus magnetic field under ZFC that we pointed
out in our paper being commented on  \cite{hmtrapping2022}. That linear behavior was modeled by the authors in the original
version of their paper \cite{minkovarxivv1} by Eq. (1) shown here. The linear fit Eq. (1) vanished from the subsequent versions of their paper Refs. \cite{talantsevarxiv,minkovtrappedpub}, where it became instead  {\it ``guides for the eye''} \cite{minkovarxivv2,minkovtrappedpub},
presumably because the authors read our paper Ref.  \cite{hmtrapping2022} and realized that doing a linear fit
Eq. (1) to ZFC data was clearly wrong. They did not acknowledge their erroroneous fit in \cite{minkovarxivv1} in the
subsequent versions of the paper \cite{minkovarxivv2,minkovtrappedpub}, nor the fact that we had
pointed it out, instead they continue to claim to this date that
ours is {\it ``the wrong model''} \cite{comment,talantsevarxiv} and to knowingly make false statements about our paper. 

We suggest that it is more important to measure and publish 
additional data of trapped moment versus magnetic field beyond the single curve for ZFC published in 2022, to show whether or not  the measurements are 
reproducible, reliable, and consistent with each other.

 \appendix*  
   \section{Letter sent to Comment authors on December 8, 2023}

  As the reader can see in Fig. 3, we clearly and explicitly told the authors
  on December 8, 2023, that their statements about our model and computer code in \cite{talantsevarxivv1}, which are the same as in 
  this Comment \cite{comment}, are wrong and misleading. Nevertheless, Ref. \cite{talantsevarxiv} remains posted in arxiv and the authors  repeat
  the same statements  one year  later
  in their Comment \cite{comment}.
         \begin{figure*} []
  \resizebox{18.5cm}{!}{\includegraphics[width=6cm]{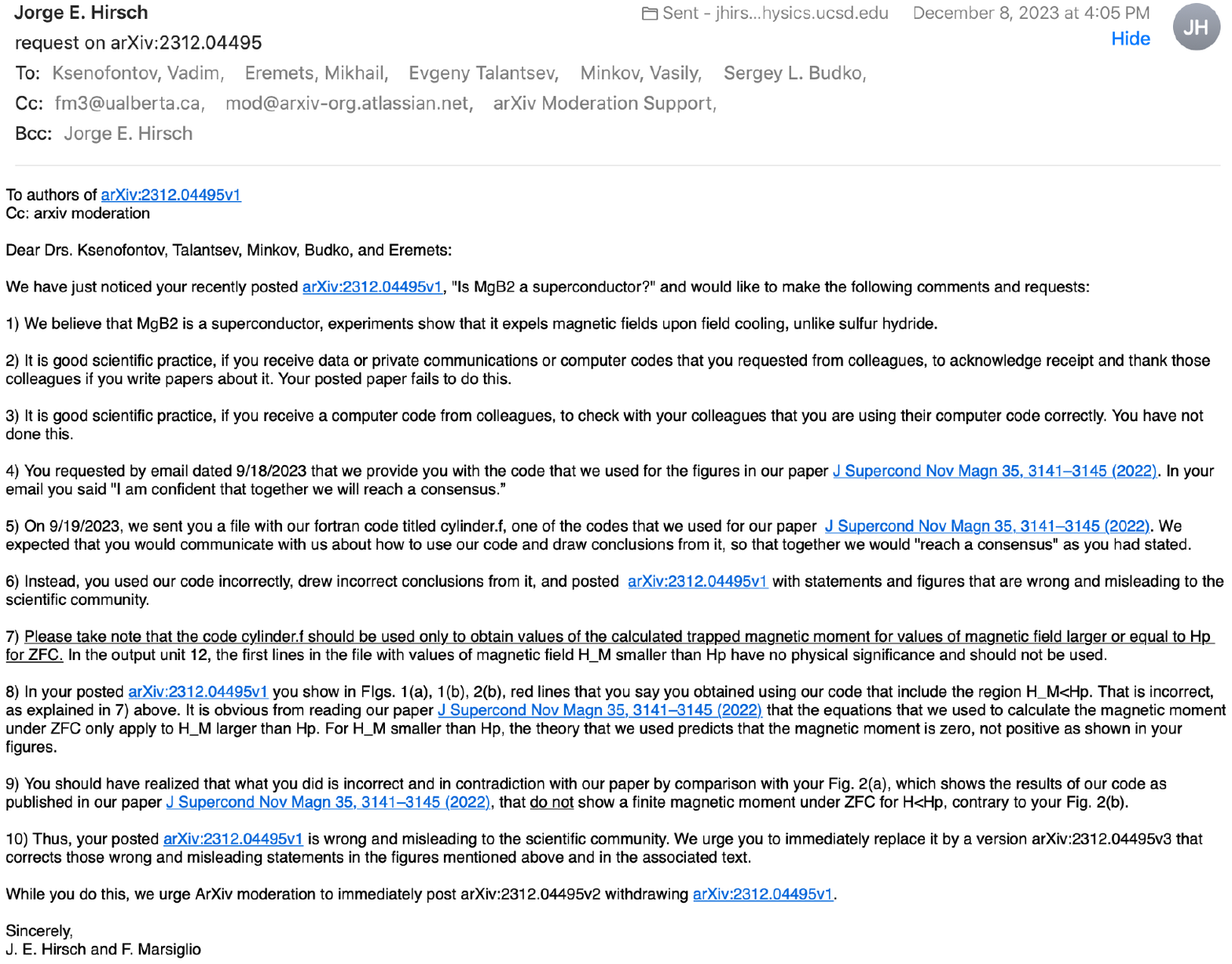}} 
 \caption {Image of email sent to authors of the Comment on December 8, 2023. }
 \label{figure1}
 \end{figure*} 

 \end{document}